\documentstyle[11pt]{article}
\input epsf
\epsfverbosetrue
\begin{document}

\renewcommand{\thefootnote}{\alph{footnote}}
\begin{titlepage}
\begin{tabbing}
\hspace{11cm} \= HIP -- 1999 -- 04 / TH \\
\> NORDITA--99/7 \\
\> \today
\end{tabbing}

\begin{center}
\vfill
{ \Large\bf A variational fit to the lattice energy of two heavy-light mesons}

\vspace{0.5cm}
A.M.~Green\footnotemark[1]$^,$\footnotemark[3],
J.~Koponen\footnotemark[1]$^,$\footnotemark[4],
and P.~Pennanen\footnotemark[2]$^,$\footnotemark[5]\\

$\,^a$Department of Physics and Helsinki
Institute of Physics, P.O. Box 9, FIN--00014 University of Helsinki, Finland \\
$\,^b$Nordita, Blegdamsvej 17, 2100 Copenhagen \O, Denmark
\end{center}
\setcounter{footnote}{3}
\footnotetext{email: {\tt anthony.green@helsinki.fi}}
\setcounter{footnote}{4}
\footnotetext{email: {\tt jmkopone@rock.helsinki.fi}}
\setcounter{footnote}{5}
\footnotetext{email: {\tt petrus@hip.fi}}
\setcounter{footnote}{0}

\vspace{1.5cm}

\date{\today}

\begin{abstract}
Recent lattice calculations on the interaction energy of two heavy-light mesons
($Q^2\bar{q}^2$) in SU(3) are interpreted in terms of the potential for the corresponding
 single heavy-light meson ($Q\bar{q}$). This model supports earlier
work, with four static quarks $Q^4$ in SU(2), that there is  a large
overestimate of the binding compared with the lattice data -- unless the basic
$Q\bar{q}$ potentials are modified by  a four-quark form factor. 
\end{abstract}
PACS numbers: 24.85.+p, 13.75.Lb, 11.15.Ha, 12.38.Gc \hfil\break

%\end{center}
\end{titlepage}

\section{Introduction}
In many-body systems, such as nuclei, involving nucleons interacting
through mesons,  it seems  justified to replace the explicit
presence of the mesons by internucleon potentials. These are
mainly between pairs of nucleons with three-nucleon potentials, in
general, playing a minor role. However, in many-body systems involving
quarks interacting through gluons, it is not at all clear whether or not  a
similar potential approach is meaningful. There are basically two
schools of thought. On the one hand, for particle physicists, the use of
only two-quark potentials in multiquark systems is not even a discussion
point, since they simply believe that it can not be correct due to the 
non-Abelian nature of the gluon fields that lead to gluon-gluon interactions.
In fact, others go even further by saying that the whole concept of 
potentials is not useful in these systems. In spite of these very basic 
objections,  there is a second school of thought -- mainly that of
many-body physicists well versed in  multi-nucleon problems -- that 
continue to treat multi-quark systems with standard many-body techniques
using two-quark potentials. 
One of the most recent and extensive sets of calculations
in this latter approach is to be found in Refs.~\cite{Barnes1,Barnes2}.
These describe meson-meson scattering in terms of four quarks~\cite{Barnes1}
and baryon-baryon scattering in terms of six quarks \cite{Barnes2}.
In each case two (or three) quarks are confined into a cluster by means of
an oscillator potential and then two such clusters interact via
two-quark potentials. However, there is no a priori justification for this.
Over the last few years
attempts have been made to clarify this situation by comparing the
{\em exact} energies of four-quark systems -- as calculated on a lattice
-- with standard many-body models using only two-quark 
potentials -- see Refs.~\cite{en, model1} (and references therein).
 Four-quark systems were chosen for this comparison
for several reasons:\\
1) Such systems exhibit a feature not present in $q\bar{q}$ and $qqq$ 
descriptions of mesons and baryons -- namely -- the possibility of there
being two two-quark color singlet clusters. This situation arises in 
meson-meson scattering, where the intercluster distance is not
restricted by confinement.\\
2) For systems with many quarks the ability to perform accurate lattice 
simulations decreases rapidly with the number of quarks -- four being
essentially the present day limit.\\
3) For the four-quark system, many-body models in terms of interquark
potentials can be expressed in simple and transparent forms.\\
In Refs.~\cite{en} - \cite{gmp} this comparison program has already been
made with the outcome, shown in Figure~1, that the resulting four-quark
binding energies are grossly overestimated by the models, if only
standard two-quark potentials are used. This is nothing more than the
well known van der Waals effect~\cite{waals} and, as shown in Refs.~\cite{en} --
\cite{gmp}, can be overcome by introducing a four-quark form factor 
containing a single free parameter.
In Ref.~\cite{model1} it was further shown that this single parameter was
capable of giving a reasonable understanding of 100 pieces of data --
the ground and first excited states of configurations, 
calculated on a $16^3\times 32$ lattice, from  different
four-quark geometries -- rectangles, quadrilaterals and tetrahedra -- in
addition to the squares of Figure~1. But in all cases the model
without the form factor yielded far too much binding for the ground state. 
 This concept of a form factor is a familiar and successful technique when
dealing with 
systems of interacting baryons, where the baryon-baryon potentials due to
the exchange of well known mesons (e.g. the $\pi$ or $\rho$) have their 
short-range singular behaviors modified by form factors. The form of the
latter are usually determined by first fitting  basic two-body 
baryon-baryon data before embarking on the study of multi-nucleon systems.
Here the same philosophy is attempted by first determining a four-quark
form factor from the four-quark lattice data before embarking on the
study of the multi-quark systems involved in a microscopic description
of meson-baryon or baryon-baryon scattering.  
It should be added that also in the context of forces between
quarks the idea of introducing many-body interactions
to remove the van der
Waals problem is not new. For example, it is discussed by the authors of
Refs.~\cite{Yaz, flip} when they introduce their alternative model --
commonly called the "flip-flop" model.

Unfortunately, the comparisons between the lattice data and the
proposed model  just discussed  had several shortcomings:\\
a) All four quarks were of infinite mass (i.e. the $Q^4$ static
approximation ).\\
b) The lattice calculations ignored the possibility for the
creation/annihilation of quark pairs (i.e. the quenched
approximation).\\
c) Only two colors were considered for the quarks (i.e. SU(2)). \\
In  the present work, the lattice data in Refs.~\cite{cp}--\cite{pmg}  have 
 these defects partially corrected. Firstly, only two
of the four quarks are now static with the other two having masses
comparable to that of the strange quark. This $Q^2\bar{q}^2$ system is often
referred to as that of two interacting heavy-light mesons.  
Secondly, the three color group [SU(3)] of QCD is incorporated. Finally,
there is now preliminary lattice data~\cite{UKQCD}  
utilising gauge field configurations generated with dynamical
fermions~\cite{Allton} i.e. the quenched approximation is no longer necessary.
These are major improvements in the lattice work. However, as will be
seen later, in the corresponding many-body model  some approximations
need to be made. But these are not expected to change the model predictions
qualitatively. The outcome is that the standard many-body model without
a form factor again overestimates the four-quark binding energy. 
On its own, this result from the $Q^2\bar{q}^2$ system would not be very convincing.
However, when combined
with the earlier work on the $Q^4$ system, the present result will  be seen to clearly
support the same conclusion that the use of two-quark potentials alone is 
not justified. On the positive side, the model also offers an effective
interaction -- albeit with a four-quark form factor -- that has been
tuned to fit the $Q^2\bar{q}^2$ system. The hope is now that this effective  interaction
 is more appropriate than the one without the form factor in other
multiquark systems  as encountered in meson-baryon or baryon-baryon scattering.
Of course, this is only a hope that such an effective interaction is in anyway
universal. It can only be substantiated by detailed calculations in other
multi-quark systems. At present, our main point is that this effective interaction
at least succeeds (by arrangement) in the four-quark case -- a feature
that is not true with the use of only two-quark potentials.

The above gives a purely theoretical motivation for this work. However,
a second motivation is that the $Q^2\bar{q}^2$ system being studied has
many features in common with the interaction between two $B$-mesons, the
latter being a combination of a $\bar{b}$-quark and an $u$ or $d$-antiquark.
These mesons are now fashionable, since the advent of $B$-meson factories at SLAC and
KEK has increased interest
in the structure of -- and interaction between -- $B$-mesons.
These facilities will not be able to study directly $BB$ reactions.
However, the related $B\bar{B}$ system is accessible as a final state of,
for example, the decay of the $\Upsilon(4S, 10580\ {\rm MeV})$, whose main
branching ($\geq 96\%$) is into
$B\bar{B}$. At present, the experimental emphasis with the $\Upsilon(4S)$  is the
study of the non-$B\bar{B}$ decays -- see for example~\cite{joe}.

The following model estimate of the four-quark binding energy (i.e. for 
$Q^2\bar{q}^2$)
is in two parts:\\
a) The energy $E(2)$ of the two separate  $Q\bar{q}$ systems.\\
b) The energy $E(4)$ of the complete four-quark system.\\
The binding energy [$B(4)$] of the latter is then defined to be
\begin{equation}
\label{be} 
B(4)=E(4)-2 E(2)
\end{equation}
 with both $E(2)$ and $E(4)$ requiring separate variational
calculations.

In Section 2 the energies of the heavy-light ($Q\bar{q}$) system,
extracted from the lattice calculation of Ref.~\cite{cp},
are fitted using a variational solution of the non-relativistic two-body
Schroedinger equation. 
This enables a determination to be made of an effective light 
quark mass ($m_q$) that is appropriate for the subsequent
non-relativistic four quark studies.
In Section 3 the formalism for calculating the binding energy of
the four-quark system by a second variational calculation is described and,
 in Section 4, this is compared with the lattice data of 
Refs.~\cite{cp,pmg}. Section 5 contains conclusions.

\section{The two-quark system}
In Ref.~\cite{cp} a new method was introduced for generating quark
propagators. For heavy-light ($Q\bar{q}$) mesons  this led to a considerable
improvement over earlier methods using iterative inversion and enabled
estimates to be made of the energies of states with orbital angular momentum
$L=0,1,2,3$. In addition, the splitting between the $j=L\pm \frac{1}{2}$
states could also be observed. However, in the present work we are not 
interested in this later refinement, since we average over spins.
This spin-averaged data $E(2,L,{\rm lattice})$ is now fitted with the 
non-relativistic Schroedinger equation
\begin{equation}
\label{sch2}
[T(2)+V(2)-E(2,L)]\phi=
\left[-\frac{d^2}{2m_qdr^2}+\frac{L(L+1)}{r^2}+V(2,r)-E(2,L)\right]
\phi=0.
\end{equation}
Here there are two unknowns -- the effective quark mass $m_q$ and the
interquark potential $V(2,r)$. 
Since the two-quark potential will be an important ingredient in the
four-quark model, care must be taken in choosing one that is appropriate
in the sense that it is most realistic over that range of $r$ dominant in the
four-quark problem. 
Later it will be seen that it is necessary to perform spacial
integrations over $V(2,r)$ and that the integrands are peaked in the range
of 2 to 4 lattice spacings. 
We, therefore, generate a two-quark potential with
 40 configurations on a $16^3\times 24$ lattice at $\beta=5.7$ using two
fuzzing levels (2,13). The basic data is given at
the on-axis points at $r/a=2,...9$ and is well fitted
($\chi^2$/dof=1.65) by the form (in fm$^{-1}$ and $r$ in fm)
\begin{equation}
\label{v2fit}
aV(2,r)=-\frac{0.309(38)}{ r/a}+0.1649(36)r/a+0.629(25).
\end{equation}
This gives a string energy  of (445 MeV)$^{2}$ for $a=0.18$ fm.
In Table 1 the lattice data is compared with this fitted potential in
Eq.~\ref{v2fit}. There it is seen that the fit is good over the
important
range of $r\sim (2 - 3)a$. In Ref.~\cite{edwards} an alternate form of
$V(2,r)$ is given. However,
this is designed to get a good fit to the lattice data for large values
of $r$ in order to extract  an estimate of the string energy, whereas
we are more interested in values of $r$ only upto about $5a$. 
Using the same form of parametrization as in Eq.~\ref{v2fit} their
corresponding parameters are 0.2618 (i.e. $\pi /12$), 0.1504 and 0.6674
respectively. This results in a string energy  of (425 MeV)$^{2}$ --
slightly 
less than our value. The greatest difference between the two forms
is the additive constant, so that at $r=1$  we have $V(2)=-138$ MeV,
whereas they have --179 MeV. This is in spite of the fact that, in both
cases,
 at  $r=1$ the original lattice data are well within $1\%$ of each
other.  
However, it is expected that our potential is more realistic at such
small 
values of $r$, since we introduce an extra degree of freedom by allowing
the
strength of the coulomb term to be a free parameter -- unlike
Ref.~\cite{edwards} where it is frozen at $\pi /12$. 
Here we only need $V(2,r)$  for the quenched approximation as no fit is
performed below to the dynamical fermion data -- the reason being that
not all the necessary observables have yet been measured.

The other unknown in Eq.~\ref{sch2} is the quark mass ($m_q$) and this is
treated as a free parameter adjusted so that the $E(2,L=0,1,2,3)$ model
energies are an average fit to the corresponding lattice data.

An estimate of $E(2,L)$ from Eq.~\ref{sch2} is extracted using the
variational principle  by simply minimizing the expression
\begin{equation}
\label{E2}
\langle\phi(r)|T(2)+V(2)|\phi(r)\rangle/\langle\phi(r)|\phi(r)\rangle
\end{equation}
with  a variational wavefunction  of the form
\begin{equation}
\label{wf2}
\phi(r)=\sum^{N_2}_{i=1}\beta_i\exp(-\alpha_i r^2/2).
\end{equation}
Here the $\alpha_i$ and  $\beta_i$ are the variational parameters, but with 
$\beta_1$ fixed at unity to set the overall normalisation. Later, it will
be seen  that $N_2$, the number of terms in the sum, need not be greater 
than three to get sufficient accuracy for $E(2,L)$. 

The outcome is $m_q=400$ MeV gives a good overall fit to the
data.
At this stage no attempt is made to optimize $m_q$.
However, this value of $m_q$ does present a problem, since it is
sufficiently small that  
relativistic effects would be expected to be important. We return to
this point later.

The above  energies $E(2,L)$, and the four-quark energies  calculated in the
next section, are obtained using a  variational
procedure and so are only upperbounds. Therefore, since the binding energy 
defined in Eq.~\ref{be} can lead to a delicate cancellation between
$E(4)$ and $2E(2)$, it is necessary to know how accurate are the
estimates of these separate quantities.

To check the accuracy of $E(2)$, the method described by
Eqs.~\ref{E2}, \ref{wf2} is used to calculate the energies for the form of 
the $L$-wave Schroedinger equation treated in Ref.~\cite{eichten}.
There the eigenvalues  are given to seven
significant figures for a series of potentials of the type given in
Eq.~\ref{v2fit}. We find that  our variational estimates 
using Eq.~\ref{wf2} for $N_2=2$ or 3 are in agreement with the
exact result to about four significant figures -- an accuracy much
better than is actually needed here.

\section{The four-quark system}
 
In Ref.~\cite{model1} a model was developed for understanding the lattice
energies of four static quarks \\
$Q({\bf r_1})Q({\bf r_2})\bar{Q}({\bf r_3})\bar{Q}({\bf r_4})$
in terms of two-quark potentials. This model, in its simplest form, was 
constructed in terms of the two basis states that can be made by
partitioning the four quarks into two color singlets - namely -
\begin{equation}
\label{AB}
A=[Q(1)\bar{Q}(3)][Q(2)\bar{Q}(4)] \ {\rm and} \ 
B=[Q(1)\bar{Q}(4)][Q(2)\bar{Q}(3)], 
\end{equation}
where $[...]$ denotes a color singlet.
These two states are not orthogonal and have a normalisation matrix of
the form -- see Ref.~\cite{masud}.

\begin{equation}
\label{Nf}
\bf{N}(f)=\left( \begin{array}{ll}
 1   & \frac{1}{3}f \\
\frac{1}{3}f & 1\\
\end{array} \right).
\end{equation}
In the extreme weak coupling limit the parameter $f=1$ and in the strong
coupling limit $f=0$. However, for intermediate situations 
it is parametrised as
\begin{equation}
\label{fexp}
f({\bf r_1},{\bf r_2},{\bf r_3},{\bf r_4})
=\exp[-b_s k_f^S 
S({\bf r_1},{\bf r_2},{\bf r_3},{\bf r_4})],
\end{equation}
where $b_s$ is the string energy,
$S(\bf{r_1},\bf{r_2},\bf{r_3},\bf{r_4})$ is an area defined by the
positions of the quarks and $k_f^S$ is a free parameter. 
As discussed in the Introduction, a single value of $\approx 0.5$ for $k_f^S$ was
capable of giving a reasonable understanding of 100 pieces of data --
the ground and first excited states of configurations from six different
four-quark geometries calculated on a $16^3\times 32$ lattice.
In this model the interaction between the quarks is expressed as a
potential matrix of the form
\begin{equation}
\label{Vf}
\bf{V}(f)=\left( \begin{array}{ll}
 v(13)+v(24)   & V_{AB} \\
V_{AB} & v(14)+v(23) \\
\end{array} \right),
\end{equation}
where
$V_{AB}=-\frac{f}{3}\left[v(13)+v(24)+v(14)+v(23)-v(34)-v(12)\right]$ 
as expected in the weak coupling limit with the
one-gluon-exchange-potential
\begin{equation}
\label{OGE}
V=-\frac{1}{3}\sum_{i\leq j}\lambda_i \lambda_jv_{ij} \ \ {\rm and } \ \
v_{ij}=-\frac{e}{r_{ij}}.
\end{equation}
Away from the weak coupling limit, $f$ is no longer unity and in addition
$v_{ij}$ is taken to be the full two quark potential of Eq.~\ref{v2fit}.
The  energy of the four static quarks is then given by diagonalising
\begin{equation}
\label{VN}
|{\bf V}(k_f)-E(4,{\rm static},k_f){\bf N}(k_f)|\psi=0.
\end{equation} 
This model, although very simple, contains the same basic
assumptions made in the more elaborate many-body models that incorporate
kinetic energy e.g. the Resonating Group Method\cite{Yaz}.
It is, therefore, reasonable that this simplified model can to some
extent check the validity of its more elaborate counterparts.

When only two of the four quarks are static the corresponding
matrices for\\
$Q({\bf r_1})Q({\bf r_2})\bar{q}({\bf r_3})\bar{q}({\bf r_4})$ can be
expressed in a similar form but where the matrix elements are now
integrals over the positions of the two light antiquarks. Below we
consider basis state $A$ to be the one realised as two separate
heavy-light mesons -- $[Q(1)\bar{q}(3)]$ and $[Q(2)\bar{q}(4)]$ -- when the 
distance ${\bf R}={\bf r_1}-{\bf r_2}$ between
the two heavy quarks becomes large. In this state the convenient
coordinates are then 
${\bf s_1}={\bf r_3}-{\bf r_1}$ and ${\bf s_2}={\bf r_4}-{\bf r_2}$, whereas
for the other partition $B$ the convenient
coordinates are  
${\bf t_1}={\bf r_3}-{\bf r_2}={\bf s_1}+{\bf R}$ and ${\bf t_2}
={\bf r_4}-{\bf r_1}={\bf s_2}-{\bf R}.$
 We also use the definition ${\bf u}={\bf r_3}-{\bf r_4}.$

The variational wavefunction is now taken to have the form
\begin{equation}
\label{wf4}
\psi({\bf r_i},f)=f^{1/2}({\bf r_1},{\bf r_2},{\bf r_3},{\bf r_4})
\sum^{N_4}_{i=1}\exp(-\tilde{\bf{X}}{\bf M}_i\bf{X}),
\end{equation}
where $\tilde{\bf{X}}=(\bf{s_1}, \  \bf{s_2}, \   \bf{R})$ and each matrix
$\bf{M}_i$ has the form
\begin{equation}
\label{M}
  {\bf M}_i = \frac{1}{2}\left( \begin{array}{lll}
a_i&b_i&c_i\\
b_i&d_i&e_i\\
c_i&e_i&g_i\\
\end{array} \right).
\end{equation}
Since the present problem considers the masses of the
light quarks to be equal, we in fact use a simplified form of
$\bf{M}_i$ with $b_i=0, \ d_i=a_i$ and $e_i=c_i$. This is not necessary,
but it is expected to be the dominant term in such a symmetric case.
Already for $N_4=2$, this wavefunction is indeed 
adequate for giving sufficiently accurate  four-quark binding energies.
Even this choice involves five free parameters $(a_1, \ c_1, \ a_2, \
c_2, \ g_2)$ in the variation -- with 
$g_1$ being fixed at unity to set the overall normalisation.
In Eq.~\ref{M} the parameters $a_i$ are analogous to the $\alpha_i$
in Eq.~\ref{wf2} and the $g_i$ play the role of the $\beta_i$.
In what follows the positions of the
light quarks are integrated over leaving matrix elements that are
functions of {\bf R}. In order to achieve this in any practical way it
is necessary to have a form for
$f({\bf r_1},{\bf r_2},{\bf r_3},{\bf r_4})$ that has a simpler spatial
dependence than the area $S$  used earlier   in  Eq.~\ref{fexp}  for the 
four static quark case. Here we take the very symmetric form advocated in 
Ref.~\cite{masud} -- namely --
\begin{equation}
\label{fsym}
f=\exp\left[-k_f b_s \sum_{i\leq j}r_{ij}^2 \right].
\end{equation}   
It should be emphasised that this form of $f$ is purely for numerical 
simplicity leading to analytical expressions for all matrix elements.
Again $k_f$ is a free parameter, which should be adjusted to fit the
four-quark lattice energies.

The wavefunction in Eq.~\ref{wf4} is used for both states $A$ and $B$.
This is an approximation that appears to work well for the 
$Q^2\bar{q}^2$ system, since $A$ and $B$ are similar in structure for
the $R$ values of interest here.

The normalisation matrix can now be written as
\begin{equation}
\label{Nfs}
{\bf N'}(R, \ k_f)=\left( \begin{array}{ll}
 N(R, \ 0)   & \frac{1}{3}N(R, \ k_f) \\
\frac{1}{3}N(R, \ k_f) & N(R, \ 0)\\
\end{array} \right),
\end{equation}
where --  after integrating over $\bf{s_1}, \  \bf{s_2}$ --
$N(R, \ k_f)$ can be expressed as a sum of  terms of the form 
\begin{equation}
\label{Nfi}
\frac{\pi^3}{(aX)^{3/2}}\exp\left[-(Z-\frac{Y^2}{X})R^2\right],
\end{equation}
where $a=0.5(a_i+a_j)+3k_f$, $c=0.5(c_i\pm c_j)+ 2k_f$, 
$d=0.5(c_i\pm c_j)- 2k_f$, 
$g=0.5(g_i+g_j)+4k_f$, $X=a-k_f^2/a$, $Y=c+k_fd/a$ and $Z=g-d^2/a$.

Since two of the quarks are not static there is now also a kinetic energy
matrix
\begin{equation}
\label{Kfs}
{\bf K'}(R, \ k_f)=\left( \begin{array}{ll}
 K_3(R,0)+K_4(R,0)   & \frac{1}{3}[K_3(R,k_f)+K_4(R,k_f)] \\
\frac{1}{3}[K_3(R,k_f)+K_4(R,k_f)] &  K_3(R,0)+K_4(R,0)\\
\end{array} \right),
\end{equation}
where, for example,
\begin{equation}
\label{Kfsp}
 K_3(R,k_f)=\int d^3s_1d^3s_2 \psi^{\star}(k_f)\left[-\frac{d^2}
{2m_q dr_3^2}\right]\psi(k_f). 
\end{equation}
Again these integrals can be expressed in forms similar to that in
Eq.~\ref{Nfi}.

Finally, the  potential matrix has the form
\begin{equation}
\label{Vfs}
{\bf V'}(R, \ k_f)=\left( \begin{array}{ll}
 \langle v(13),0\rangle+\langle v(24),0\rangle   & \langle V_{AB},k_f\rangle \\
\langle V_{AB},k_f\rangle & \langle v(14), 0\rangle+\langle v(23), 0\rangle \\
\end{array} \right),
\end{equation}
where
\[\langle V_{AB},k_f\rangle =\]
\begin{equation}
\label{Vfsp}
\frac{1}{3}\left[\langle v(13),k_f\rangle+
\langle v(24),k_f\rangle+\langle v(14),k_f\rangle+\langle v(23),k_f\rangle
-\langle v(34),k_f\rangle-N(R,k_f) V(2,R)\right]. 
\end{equation} 
Here $N(R,k_f)$ is defined in Eq.~\ref{Nfs}, $V(2,R)$ is the 
 potential between the two heavy quarks and,
for example,
\begin{equation}
\label{v13}
\langle v(13),k_f\rangle= \int d^3s_1d^3s_2
\psi^{\star}(k_f)V(s_1) \psi(k_f).
\end{equation}
For potentials of the form in Eq.~\ref{v2fit}, these integrals can be 
expressed in terms of Error functions.
The energy $E(4,k_f)$ of the two heavy-light meson system is then obtained by 
diagonalising
\begin{equation}
\label{KVN}
|{\bf K}'(R, k_f)+{\bf V}'(R, k_f)-E(4,R, k_f){\bf N}'(R, k_f)|\psi=0.
\end{equation} 
Since this is a $2\times 2$ determinant a prediction could also be made for an
excited state $E^*(4,k_f)$ and the corresponding binding energy $B^*(4).$

We saw that the variational method worked very well for the two-quark
energies. In the four-quark case,
 when the intermeson interaction $\langle V_{AB}, k_f \rangle$ is set to zero 
(i.e. $k_f\rightarrow \infty $ -- the strong coupling limit) , a necessary
condition is that $E(4,k_f=\infty)-2E(2,L=0)$ should be approximately zero.
This is found to be sufficiently 
well satisfied, provided $N_4$ in Eq.~\ref{wf4} is at least 2. There is a
small remaining repulsion of about 5 MeV  due to inadequacies in the 
$\psi(r,f)$ of Eq.~\ref{wf4} and this could presumably be made smaller by
improving this wavefunction. However,  5 MeV should be compared with
the two body energy of 709 MeV, which is made up from a kinetic energy
 of 339 MeV and a potential energy of 370 MeV. So we see that
the condition $E(4)=2E(2)$ is satisfied to within  $1\%$.    

\section{Results}
One of the main ingredients of the above model is the interquark
potential $V(2,r)$. This enters in three different contexts:\\
1) As $v(13), \  v(24), \ v(14), \  v(23)$ -- the $Q\bar{q}$ potential in Eq.~\ref{v2fit}.\\
2) As $v(34)$ -- a $\bar{q}\bar{q}$ potential. Here we assume this to also be of the
form in Eq.~\ref{v2fit}\\
3) As $v(12)$ -- a $QQ$ potential. This was calculated from the same
gauge configurations as the four-quark energies. In this case there was
no need to fit this with a function of $R$, since it is only ever needed
at discrete values of $R$ -- the ones for which the four-quark energies
are calculated.\\

\vspace{0.01cm}

This prescription for $V(2,r)$ is the one throughout this article.
However, to check the dependence of the following results on this
choice, several other options were considered. But in all cases 
the same qualitative conclusions emerged.

Given these two-quark potentials, then the results from
the four-quark model, described in the previous section, can now be compared
 with the lattice calculations  of Ref.~\cite{pmg}.
This model gives the binding energy of the four-quark state using 
a spin-isospin independent interaction. Therefore, in order to make a
comparison with the lattice data of Ref.~\cite{pmg},
which are dependent on the spin $(S_q)$ and isospin $(I_q)$ of the
$(\bar{q}\bar{q})$ subsystem, an averaging of this data must be made. 
This averaging could be avoided if, in addition to the spin independent
interaction of  Eq.~\ref{v2fit}, the model also contained
a spin dependent interaction. However, the later is expected to be of
short range -- in fact a delta-function from One Gluon Exchange -- and
so be less affected by any four-quark form factor. Therefore, by only
considering the spin independent contributions to the binding, we hope
to maximise the effect we are studying -- namely the need for such a form
factor. Of course, in further developments of the model this restriction
will clearly need to be lifted and both the spin-independent and
-dependent potentials included. 
To carry out the averaging we are
 guided by the weak coupling limit -- the same limit
already used in setting up the basic form of the model in Eqs.~\ref{Nf}
-\ref{VN}. As shown in the Appendix of Ref.~\cite{Wein} the two basis
states $A$ and $B$ have a color structure of the form:
\begin{equation}
\label{WI1} 
|A\rangle=|(1\bar{3})(2 \bar{4})\rangle=\sqrt{\frac{1}{3}}|[12][\bar{3}\bar{4}]\rangle
+\sqrt{\frac{2}{3}}|\{12\}\{\bar{3}\bar{4}\}\rangle
\end{equation}
and
\begin{equation}
\label{WI2} 
|B\rangle=|(1\bar{4})(2 \bar{3})\rangle=-\sqrt{\frac{1}{3}}|[12][\bar{3}\bar{4}]\rangle
+\sqrt{\frac{2}{3}}|\{12\}\{\bar{3}\bar{4}\}\rangle,
\end{equation}
where $(...)$ denotes a color singlet, $[...]$ a color triplet and 
$\{...\}$ a color sextet. The overlap $\langle A|B\rangle$ gives the
factor of 1/3 appearing in Eq.~\ref{Nf} and also we have the
relationship
\begin{equation}
\label{WI3}
|A\rangle-|B\rangle=\frac{2}{\sqrt{3}}\left([12][\bar{3}\bar{4}]\right)
\end{equation}    
i.e. in this combination of $A$ and $B$ the $(\bar{q}\bar{q})$ subsystem
is in a color triplet state and so
antisymmetric in color. Since the lattice data only involves
S-wave interactions, $(I_q, S_q)$ must be (0,0) or (1,1) to ensure
overall antisymmetry for the interchange $\bar{q}_3\leftrightarrow \bar{q}_4$.
For an interquark interaction of the form
$V=V_0+{\bf s(\bar{q}_3)}.{\bf s(\bar{q}_4)}V_s$, we
have $V(0,0)=V_0-3V_s/4$ and $V(1,1)=V_0+V_s/4$. Therefore, to extract the
effect of $V_0$ -- the spin independent part of the interaction -- we
need for $|A\rangle-|B\rangle$ the combination $V_0=[3V(1,1)+V(0,0)]/4$.
However, for $R=0$ the two diagonal matrix elements in each of the matrices 
${\bf N'}(R, \ k_f), \ {\bf K'}(R, \ k_f), \  {\bf V'}(R, \ k_f)$ 
in Eqs.~\ref{Nfs}, \ref{Kfs}, \ref{Vfs} are {\em equal}, so that from Eq.~\ref{KVN}
the wavefunction of the ground state is, indeed, simply  proportional
to $\frac{1}{\sqrt{2}}[A-B]$. As $R$ increases the amplitude of
$A$ then increases to 0.75 at $R=0.18$ fm and eventually to unity 
for $R \geq 0.5$ fm. Therefore, for the small values of $R$ of interest
here, the ground state wavefunction is automatically approximately
proportional to $[12][\bar{3}\bar{4}]$ as in Eq.\ref{WI3}. This
suggests that the above method for extracting $V_0$ should be sufficiently
accurate.

Figure~2 compares the lattice results for $V_0$ 
(with and without dynamical sea quarks) of Ref.~\cite{pmg}  using
the weak coupling model (i.e. $f=1$ in Eq.~\ref{wf4} or $k_f=0$ in 
Eq.~\ref{fsym}). The model is the one designed for the lattice data
in the quenched approximation i.e. using $V(2,R)$ of Eq.~\ref{v2fit}.
The preliminary data of Ref.~\cite{UKQCD} that includes dynamical
sea quarks is simply included to show that there are no dramatic changes in the
binding energies. The quenched data from Ref.~\cite{pmg} used the lattice
parameters $\beta=5.7,\ \kappa=0.14077,\ C_{\rm SW}=1.76$ giving a lattice
spacing of $a=0.170$ fm and mass ratio $M_{\rm PS}/M_{\rm V}=0.65$, 
whereas the unquenched data~\cite{UKQCD} has
$\beta=5.7,\ \kappa=0.1395,\ C_{\rm SW}=1.52$ and $a=0.142$ fm, 
$M_{\rm PS}/M_{\rm V}=0.72$. Figure~2 clearly shows that for $R=0.18$ fm 
the model overestimates the binding by over a factor of three. For larger
values of $R$ the relative error bars on the small energies are, at present, too
large to make any conclusions. Unfortunately, the lattice data at
the smallest values of $R$ could well still contain lattice artifacts that are
not completely cancelled in the difference defining the binding energy in
Eq.~\ref{be} -- a point discussed later.
However, when the factor $f$ is no longer unity, the model binding decreases
considerably. This is shown in Figure~3  for
$k_f=0.05, \ 0.10, \ 0.15$ and  $m_q=400$ MeV with the optimal value
$k_f\approx 0.10$ fitting the first lattice data point. 
Also for this value of $k_f$   the variation with $m_q$ is found to
be small. It should be added that the lattice spacing $a$ differs
slightly (0.18 fm vs 0.17 fm) between that used for $V(2,r)$ based on  
Ref.~\cite{cp} and the
four-quark data of Ref.~\cite{pmg}. This is due to a difference in the
procedure for extracting $a$. We could have scaled the data to a common
$a$. But this refinement is not necessary for the present stage of precision.  

The form factor in  Eq.~\ref{fsym} can also be used in Eq.~\ref{VN} to 
fit the $Q^4$ lattice data in Figure~1. There it is seen with the dashed
lines that $k_f=0.075$ gives a good fit to both the ground and excited
states over the whole range of $R$ values considered. This value of
$k_f$ is in good agreement 
 with that needed in Figure~2 -- $k_f\approx 0.10.$
 
In Figure~2 is also shown the corresponding binding energy when the
2-quark potential of Ref.~\cite{edwards} is used  in the model (dashed
line). The difference between these two  gives an estimate of the
range of values that can be expected using different 2-quark potentials.
In all cases, the model overbinds by a large factor. 
The difference is also readily understandable by considering the small
$R$
limit of the model. There it is seen that it is the $V(2,R)$ in
Eq.~\ref{Vfsp}
that dominates. This results in $E(4,R)\rightarrow V(2,R)/2$ as 
$R\rightarrow 0$. If this limit were already reached at $R=1$, then  we 
would expect -- from the estimates of $V(2,R=1)$ given 
after Eq.~\ref{v2fit} -- that the binding energies from the two versions
of $V(2,R)$ should be about --70 MeV and -- 90 MeV. Eventhough these
limits underestimate by almost a factor of two the model values in 
Figure~2, they have the correct trend and also emphasise the importance
of needing a good model for $V(2,R)$.

The above comparisons between the four-quark lattice energies and the
variational estimates have at least three shortcomings:

\vskip 0.2 cm

a) The most serious problem with the  variational calculation is that
$m_q$ is sufficiently  small that a relativistic form for the 
kinetic energy should
be used both in the two- and four-quark variational formulations for the
energy. An indication of the effect of using the non-relativistic approximation 
can be estimated by comparing the effective two-body kinetic energy in
the four quark case with the kinetic energy in the original
two-body problem i.e. compare $ K_3(R,0)$ [or $ K_4(R,0)$ -- they are equal] in
Eq.~\ref{Kfs} with
$\langle\phi(r)|T(2)|\phi(r)\rangle$ in Eq.~\ref{E2}, which has the
value 339 MeV. It is found that
 $ K_3(R,0)$ is quite dependent on $k_f$ but much less so on $R$. 
For example, at $R=0.18$ fm and $ k_f=0.0, \ 0.15, \ 0.25, \ 0.45 $ we find 
$ K_3(R,0)$=286, 328, 338, 346 MeV respectively.
 Therefore, in the difference $B(4)=E(4)-2E(2,0)$
much of the kinetic energy terms cancel -- especially after the inclusion of
the form factor needed to tune the model energy to the lattice value. 
In fact, for the optimal value of $ k_f\approx 0.25$ the cancellation is
within a few MeV. For larger values of $R$ the cancellation is even 
more complete.
This leads to the expectation (hope)
that a similar cancellation will occur in a relativistic formulation and
that the binding energies calculated with the present non-relativistic
model are indeed quite realistic. 

\vskip 0.2 cm

b) The second problem is that there is some ambiguity in the form of the
 two-quark potential $V(2,r)$ in Eq.~\ref{v2fit}, since it is a fit to
lattice $Q\bar{q}$
potentials that are only known at discrete values of $r$. This fit was
designed to ensure that $V(2,r)$ reproduced well the lattice potentials
over the range $2\leq r/a \leq 4$. The reason for this is found from
 Table~\ref{Keff}, where  
it is shown how the effective two-body potential ($VT$)
is made up from the two potential components -- linear $VL$ and
coulomb-like $VC$ -- in $V(2,r)$ of Eq.~\ref{v2fit}. These can be
compared with the corresponding two-quark contributions, namely, 
370, 524, --154 MeV respectively. 
If we now define $n$ as the ratio $-VL/VC$, then a rough estimate of the most
important range of $r$ values for the integrals is given by
\begin{equation}
\label{ri}
r_I\approx a\sqrt{\frac{ne}{b}},
\end{equation}
where $e=0.309$ and $b=0.165$ from Eq.~\ref{v2fit}. In this case 
from Table~\ref{Keff} we have $n\approx 3$,
so that $r_I \approx (2 - 3)a \approx (0.4 - 0.5 )$ fm. At such values of $r$,
lattice artifacts should be small. Therefore, Eq.~\ref{v2fit} is expected
to be valid and yield reliable values for the radial integrals. In Figure~2 
was shown the dependence of the four-quark binding energy on
 the choice of 2-quark potential. There it was seen that the potential
proposed in Ref.~\cite{edwards} gave even more overbinding than that in
Eq.~\ref{v2fit}. In order to correct this, therefore, needs a stronger
(i.e. shorter ranged) form factor with $k_f\approx 0.4$.

\vskip 0.2 cm

c) A third shortcoming in the above formulation is in the form of the 
variational wavefunction in Eq.~\ref{wf4}. This could be improved in two
ways. Firstly, the limitations to $b_i=0, \ d_i=a_i$ and $e_i=c_i$ in 
Eq.~\ref{M} for the form of the $\bf{M}_i$ could be removed. This would 
presumably remove part of the 5 MeV difference in the
$E(4,k_f=\infty)-2E(2,L=0)$ mentioned earlier. However,
secondly, Eq.~\ref{wf4} is designed to describe states that are naturally
in terms of the ${\bf s_i}$ radial coordinates defined just before 
that equation. This means that this form is optimal for state $A$ but
not state $B$, which is described naturally in terms of the 
 ${\bf t_i}$ radial coordinates. The error introduced by this can be
roughly estimated by minimizing the energy of the excited state and
treating the ground state model energy as a prediction. 
However, in practice,  this seems to be unimportant for the values of $R$ of interest
here. For example, in the strong coupling limit, when $k_f$ is very
large (140 here),  minimizing the ground state energy at $R=0.26$ fm 
leads 
to $B(4), \ B^*(4)$ having 4.8 and 108 MeV respectively, whereas the
corresponding numbers are 7.5 and 105 MeV when it is the excited state
that is targetted in the minimization. At smaller values of $R$ the
differences are even smaller.

\section{Conclusions}
The main conclusion is most clearly demonstrated in Figure~2, where it is
seen that the present
model {\em in the weak coupling limit} overestimates the binding energy given
by the lattice simulation. This means that the four quark systems
($Q^2\bar{q}^2$) studied here
{\em cannot} be described simply in terms of two-quark potentials. The
effect of the latter needs to be  suppressed and this is achieved here  
 by introducing the explicit four-quark form factor $f$ shown in
Eq.~\ref{fsym}. By itself this conclusion  would not be very 
convincing, since it all depends on the
one lattice data point at $R/a=1$.
However, when this  result is combined with
the more reliable, but less physically interesting case, of four static
quarks from our earlier work, we see that the same trend is observed and
that the combined result adds further support to  our main conclusion.
In fact this support is not only  qualitative  but also quantitative,
since the optimal values for $k_f$ in Figures~ 1 and ~3 are quite
similar -- being 0.075 for $Q^4$ and 0.10 for $Q^2\bar{q}^2.$
Furthermore, the hope is that the form factor needed to tune the model
to the lattice data is universal in the sense that it can be used in 
systems with more than four quarks.  This is the ultimate aim
of this work -- a bridge between few quarks systems amenable to
lattice calculations and multi-quark systems that are beyond such
methods and so rely on more conventional many-body techniques.

In the future both the lattice simulation and the model will be
improved. As mentioned above, the lattice data will eventually
give binding energies with smaller error bars at larger values of $R$.
This could be not only for the ground state but also for  excited
states. Also the effect of using unquenched light quark 
propagators is now becoming possible~\cite{UKQCD}. 
The model is also capable of being extended and improved in several ways.
For example, in the $Q^2\bar{q}^2$ system it could be extended to
include a spin dependence in the basic interaction of Eq.~\ref{v2fit}. 
Also the basic form of the
variational wavefunction in Eq.~\ref{wf4} could be improved by using
more of the parameters in the matrix ${\bf M}$ of Eq.~\ref{M}. However,
such an improvement would presumably lead to  more binding and so
increase even more the difference between the lattice data and the basic
two-quark potential model without the four-quark form factor. 
Of course, the point needing most improvement is the treatment of
the kinetic energy, eventhough the indications are that the kinetic
energy in the four- and two-body systems to a large extent cancel each other.
 At present, only the non-relativistic form is used.
However,  it should be
possible to treat a semirelativistic version of the model, where 
$m_q+p^2/2m_q$ is simply replaced by $\sqrt{p^2+m_q^2}$ and the kinetic
energy integrals performed in momentum space. Unfortunately, this will result
in a formulation that involves a  one-dimensional numerical integration over  
momentum.

Another
improvement that could become necessary with improved data would be the
addition of new basis states with excited two-body potentials. The excited
states were found necessary in our previous fit to a large set of 
precisely measured energies in the static SU(2) case, 
with the main conclusion about these states being that they 
have a larger effect than the ground state
potentials on the binding for large $R$ (beyond 0.5 fm). However, in the
present more dynamic system an additional effect for $R\ge 0.5$ fm would
come from meson exchange, for which impressive agreement in this distance
range was found in Ref.~\cite{pmg}. A Yukawa potential effective at 
long distances could be added to relevant matrix elements of our model in 
order to reproduce the observed deuson-like behavior~\cite{nt}. 
Such a potential
is expected to be essential for the observed binding of 
$(I_q, S_q) = (0,1); (1,0)$ states not discussed in the present work.

The authors wish to thank Ted Barnes,
Tim Klassen, Chris Michael and Jean-Marc Richard for useful discussions and
correspondence. We are also grateful to the CSC in Espoo for their excellent 
computational resources and to the Finnish Academy for  supporting 
this project.

\newpage 

\vskip 0.5 cm

{\bf Figure Captions}
\vskip 0.5 cm

\noindent Figure~1.  The case of four static quarks $(Q^4)$ in SU(2)
placed at the corners of
squares of side $R/a$, where $a\approx 0.12$ fm. This shows the binding
energy (aB) in lattice units as a function of $R/a$. 
The lattice results are for $\beta=2.4$ 
on a $16^3\times 32$ lattice with the dots(crosses) showing the
ground(excited) state energies -- with error bars.
The lower(upper) solid  curve is a model estimate for the
ground(excited) binding energy
obtained assuming only 2-quark interactions. 
This figure is similar to the one in ~\protect\cite{gmp}. 
The lower(upper) dashed  curve is the effect of the form factor in 
Eq.~\protect\ref{fsym} with $k_f=0.075$. 
\vskip 0.5 cm

\noindent Figure~2.   
Comparison between the spin independent part
$(V_0)$ of the $Q^2\bar{q}^2$ binding energies
calculated on a lattice ~\protect\cite{pmg} (solid circles -- quenched
approximation with $a=0.170$fm)/~\protect\cite{UKQCD}
(solid squares -- with dynamical fermions and $a=0.142$fm)
and the model in the weak coupling limit ($k_f=0$). The crosses,
with the solid  line to guide the eye, use the two-quark potential
in Eq.~\protect\ref{v2fit} and the stars, with the dashed line, the
two-quark potential in Ref.~\protect\cite{edwards}. The dynamical fermion 
data is not used in any fit. It is simply included to show that it is
qualitatively consistent with the quenched data but with considerably
smaller error bars.

\vskip 0.5 cm

\noindent Figure~3. Comparison between the spin independent part
$(V_0)$ of the $Q^2\bar{q}^2$ binding energies
calculated on a lattice ~\protect\cite{pmg} and using the
variational principle with 
$k_f$ =0.15 (dotted), 0.10 (solid), 0.05 (dashed) and $m_q$=400 MeV. 
Other notation as in Figure~2.
\vskip 0.5 cm

\newpage
 
\vspace{0.5cm}

\begin{table}
\begin{center}
\caption{ The comparison, as a
function of $r/a$,  between the two-quark potential $aV(2,r)$ of
Eq.~\protect\ref{v2fit} with the additive constant removed  and the
potential given by Wilson loops in Ref.~\protect\cite{pmg}.}
\vspace{1cm}
\begin{tabular}{cccccc}
\hline
$r/a$&1&2&3&4&5\\ \hline \hline
$aV(2,r)$
&--0.144   &0.175  &0.392  &0.582   &0.763\\
~\protect\cite{pmg}&--0.126(1)&0.175(2)&0.392(3)&0.586(9)
&0.783(22)\\
\end{tabular}
\label{Vlatt}
\end{center}
\end{table}   

\begin{table}
\begin{center}
\caption{The effective two-body potential ($VT$) in MeV for the four-quark
system at $R=0.18$ fm and $k_f=0.25$. 
The $VL$ and $VC$ are the linear and coulomb-like contributions to $VT.$
 }
\vspace{1cm}
\begin{tabular}{c|ccc}
&$VL$&$VC$&$VT$ \\ \hline 
$\langle V({\bf s_1})\rangle$&300&--102&198 \\
$\langle V({\bf t_1}) \rangle$&315&--97&218 \\
$\langle V({\bf u})\rangle$&423&--72&351 \\
\end{tabular}
\label{Keff}
\end{center}
\end{table}
\vspace{0.5cm}

\newpage

\begin{figure}[ht]
\label{static}   
\vspace{14cm} % was 14  vof was -30:too low
\includegraphics{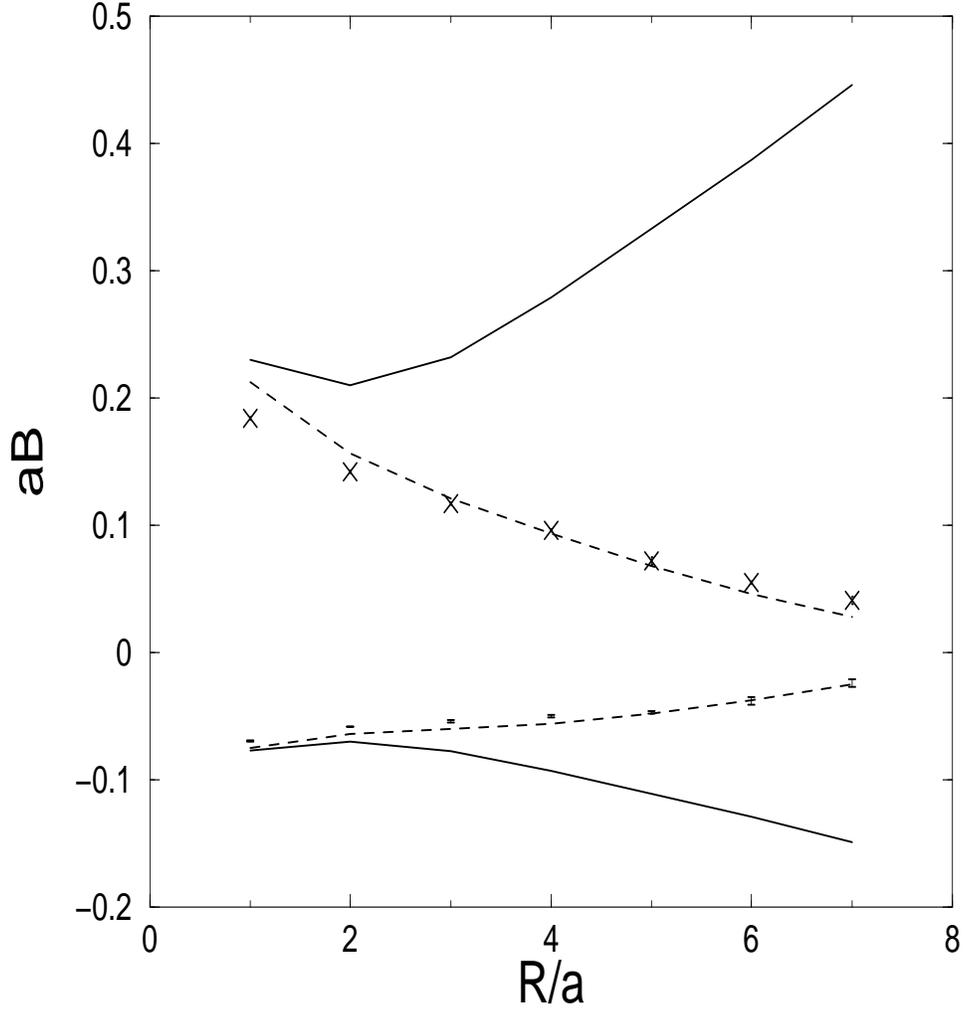}

\caption{ 
The case of four static quarks $(Q^4)$ in SU(2)
placed at the corners of
squares of side $R/a$, where $a\approx 0.12$ fm. This shows the binding
energy (aB) in lattice units as a function of $R/a$. 
The lattice results are for $\beta=2.4$ 
on a $16^3\times 32$ lattice with the dots(crosses) showing the
ground(excited) state energies -- with error bars.
The lower(upper) solid  curve is a model estimate for the
ground(excited) binding energy
obtained assuming only 2-quark interactions. 
This figure is similar to the one in ~\protect\cite{gmp}. 
The lower(upper) dashed  curve is the effect of the form factor in 
Eq.~\protect\ref{fsym} with $k_f=0.075$.
}
\end{figure}

\begin{figure}[ht]
\label{kf=0}
\vspace{14cm} % was 14  vof was -30:too low
\includegraphics{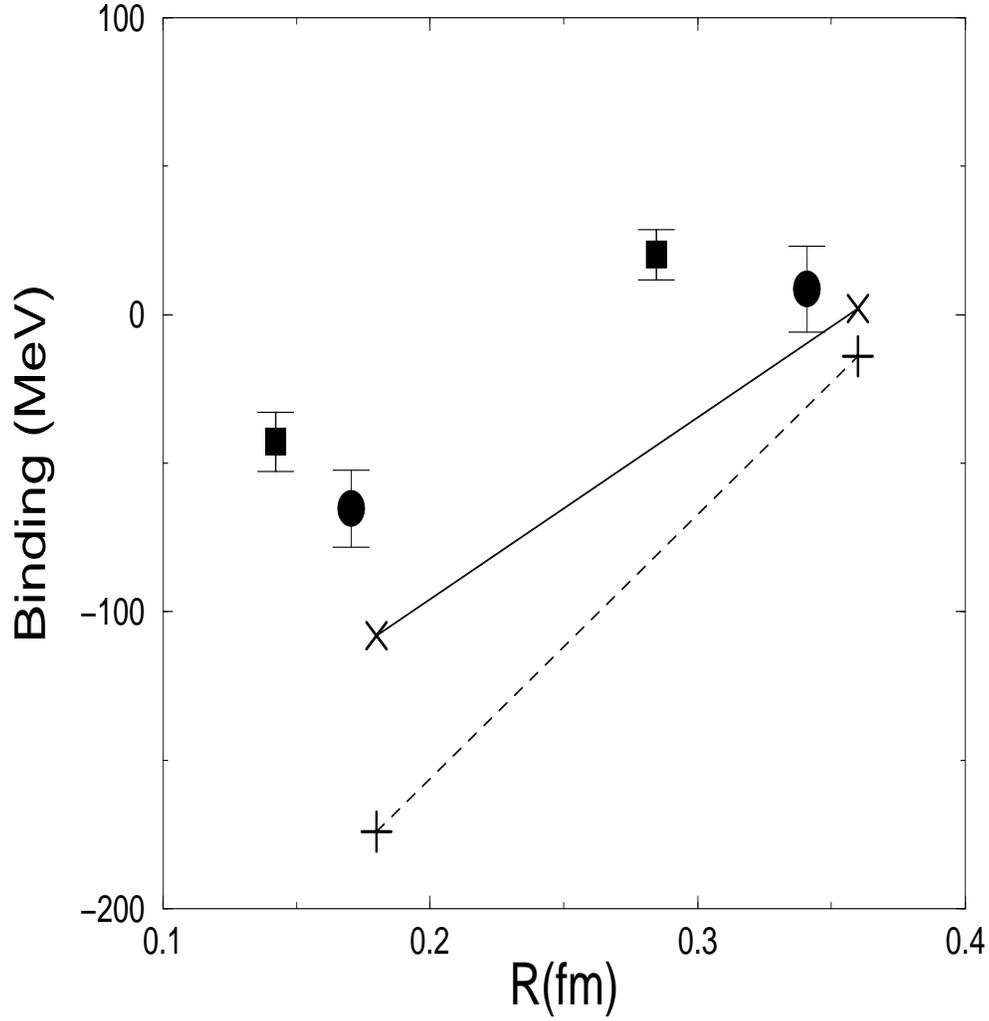}

\caption{ Comparison between the spin independent part
$(V_0)$ of the $Q^2\bar{q}^2$ binding energies
calculated on a lattice ~\protect\cite{pmg} (solid circles -- quenched
approximation with $a=0.170$fm)/~\protect\cite{UKQCD}
(solid squares -- with dynamical fermions and $a=0.142$fm)
and the model in the weak coupling limit ($k_f=0$). The crosses,
with the solid  line to guide the eye, use the two-quark potential
in Eq.~\protect\ref{v2fit} and the stars, with the dashed line, the
two-quark potential in Ref.~\protect\cite{edwards}. The dynamical fermion 
data is not used in any fit. It is simply included to show that it is
qualitatively consistent with the quenched data but with considerably
smaller error bars.  }
\end{figure}
\begin{figure}[ht]
\label{kf}
\vspace{14cm} % was 14  vof was -30:too low
\includegraphics{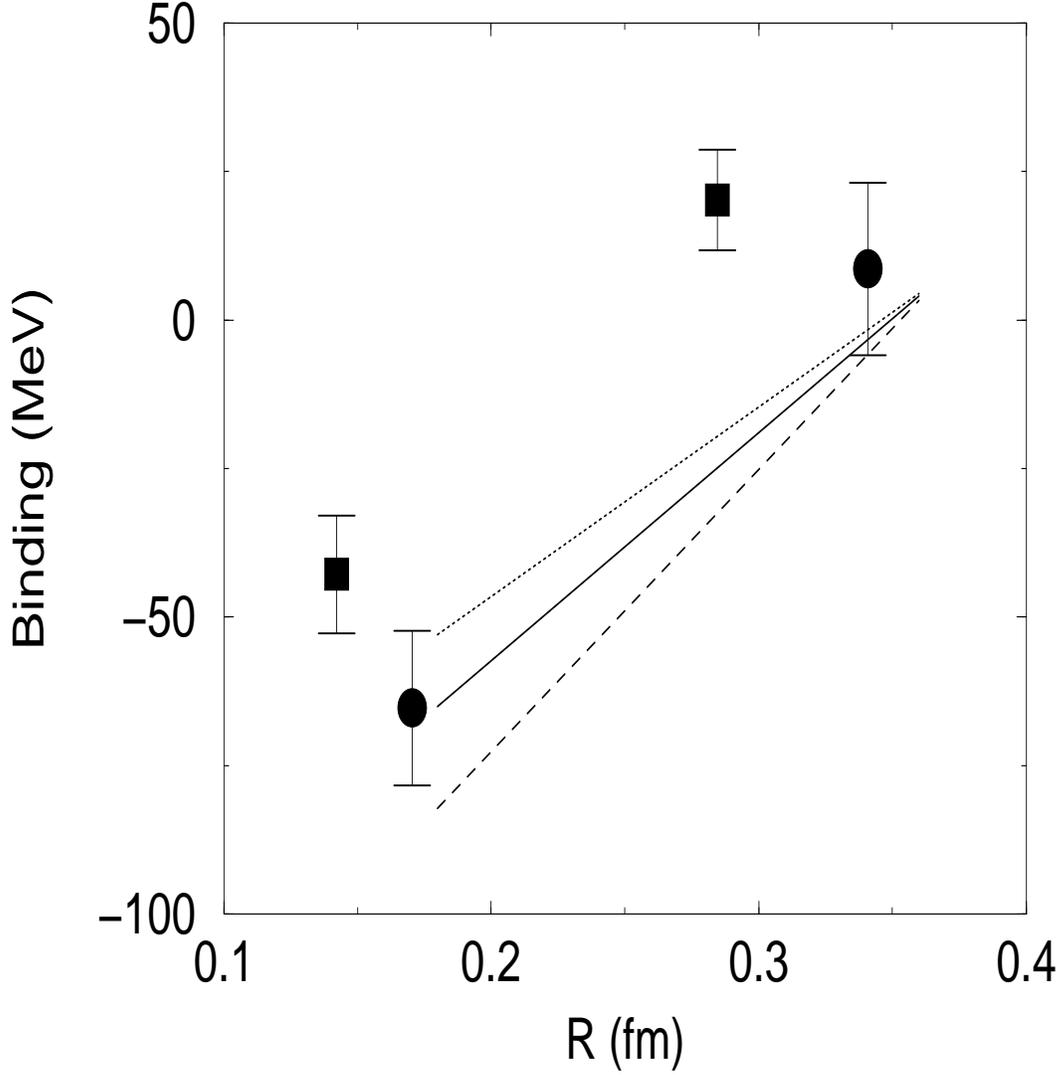}

\caption{
Comparison between the spin independent part
$(V_0)$ of the $Q^2\bar{q}^2$ binding energies
calculated on a lattice ~\protect\cite{pmg} and using the
variational principle with  
$k_f$ =0.15 (dotted), 0.10 (solid), 0.05 (dashed) and $m_q$=400 MeV. 
Other notation as in Figure~2.}
\end{figure}

\end{document}